\documentclass[12pt]{article}
\usepackage{graphicx}
\usepackage{amssymb}
\usepackage{amsthm}
\usepackage{amsmath}
\usepackage{setspace}
\newtheorem{Def}{Definition}
\newtheorem{Remark}[Def]{Remark}
\newtheorem{Ex}[Def]{Example}
\newtheorem{Theorem}[Def]{Theorem}

\newtheorem{Corollary}[Def]{Corollary}
\newtheorem{Proposition}[Def]{Proposition}

\newenvironment{Proof}{\noindent{\bf  Proof.}}{\QED\bigskip}

\newcommand{\N}{{\rm I\! N}}
\newcommand{\R}{{\rm I\! R}}

\newcommand{\QED}{\hfill $\Box$}

\begin{document}
\baselineskip=17pt

\doublespacing
\begin{center}
{\large \bf{Generalized semi-Markovian dividend discount model: risk and return}}
\end{center}
\begin{center}
{\sc Guglielmo D'Amico}\\
Department of Pharmacy, "G.d'Annunzio" University of  Chieti-Pescara\\
via dei Vestini 30, 66013 Chieti, Italy.\\
e-mail: \texttt{g.damico@unich.it}
\end{center}

\indent {\bf Abstract:} The article presents a general discrete time
dividend valuation model when the dividend growth rate is a general
continuous variable. The main assumption is that the dividend growth
rate follows a discrete time semi-Markov chain with measurable space
$(E, \mathcal{E})$. The paper furnishes sufficient conditions that
assure finiteness of fundamental prices and risks and new equations that
describe the first and second order price-dividend ratios.
Approximation methods to solve equations are provided and some new
results for semi-Markov reward processes with Borel state space are
established.\\
\indent The paper generalizes previous contributions dealing with
pricing firms on the basis of fundamentals.

\newpage

\setcounter{equation}{0} \setcounter{Def}{0}
\section{Introduction}
\label{Pref} \setcounter{equation}{0} \setcounter{Def}{0}

One of the main approaches to calculate the value of a firm is by means of fundamentals. Fundamental analysis consists in discounting of future cash flows  by using a required rate of return on the stock and in considering the value of the stock equal to the present value of this stream of payments, see e.g. Kettel (2002). The rate of return includes a risk premium for the shareholders to compensate for the uncertainty associated with the future evolution of cash flows.\\
\indent Stockholders expect to receive cash flows from the firm in term of dividends therefore, usually cash flows are represented by dividend streams. Nevertheless, it is possible to replace dividends
with other financial variables like sales (see, e.g. Damodaran 1994)
or earnings and payout ratios (see e.g. Sharpe and Alexander 1990). Since generally, stockholder receives only dividends from the firm the majority of the models are based on dividends, for this reason henceforth, we will assume that the firm pays dividends.\\
\indent Almost all studies in literature impose sufficient structures
on the dividend growth variable to permit computable expressions of the present value of future dividends. The seminal paper by Gordon and Shapiro (1956) considers a constant dividend growth rate. Many variants of the Gordon and Shapiro model have been suggested in literature. These variants impose ever less stringent assumptions on the dividend process. For example, the papers by Brooks and Helms (1990) and Barsky and DeLong (1993) consider multistage models with dividend growth rates changing deterministically among the stages. Models based on Markov chains were proposed in Hurley and Johnson (1994;1998), Yao (1997) and Ghezzi and Piccardi (2003) and, in general, regime switching in the dividend process were advanced in Gutierrez and Vasquez (2004). The results of these articles were encompassed into a semi-Markovian framework as provided by D'Amico (2013) where the semi-Markov hypothesis was advanced and validated.
Mainly due to their generality and flexibility, recently semi-Markov processes have acquired even more interest in many areas of financial modeling ranging from credit rating dynamics (see, e.g. D'Amico et al. (2005), Vasileiou and Vassiliou (2006),  Vassiliou and Vasileiou (2013) and Vassiliou (2014)) to high frequency finance (see D'Amico and Petroni (2011) and D'Amico and Petroni (2012)).

More recent contributions on dividend valuation models rely on one side on the adoption of Geometric and Additive Bernoulli Processes where the dividend process assumes a continuous set of values, see Hurley (2013) and on the other side on the derivation of formulas for the variance of the fundamental prices obtained in a binomial based model, see Agosto and Moretto (2013).

\indent These two recent contributions necessitate a unifying treatment under the adoption of more general stochastic models that overcome the strong simplifying assumption of Bernoulli process. Indeed, it is necessary to avoid imposing strong hypotheses on the dividend process in favor of more general models that let data to speak for themselves without imposing stringent assumptions.

\indent Therefore the aim of this paper is to generalize all these
contributions by assuming that the growth dividend rate is a discrete time semi-Markov chain with Borel state space. Then, we advance a more formal and abstract dividend valuation model that contains as particular cases all the previous cite works, D'Amico (2013) included. 

The second feature of this study is that we derive sufficient conditions that assure the finiteness of fundamental prices and that satisfy the transversality condition that avoid the presence of speculative bubbles within this general semi-Markov environment. Moreover we extend these results by computing the second order
moment of the price process which reveals to be useful for measuring the riskiness of the stock. To do this we determine the fundamental formula of the risk (second order moment of the price process) and we describe methods for computing it.

\indent It is important to remark on the presence of contributions that generalize the Gordon and Shapiro model by allowing arbitrary dividend growth and discount rate processes, see Donaldson and Kamstra (1996). The Donaldson and Kamstra's procedure involves a Monte Carlo simulation and numerical integration of the possible paths followed by the joint processes of dividend growth and discount rates.

\indent The paper is organized as follows: first, in Section 2, we give minimal definitions on discrete time semi-Markov chains with Borel state space. Next, Section 3 presents generalities on the dividend valuation model, on the semi-Markov stock model and the relevant results of the proposed dividend valuation model. Continuing, Section 4 presents some concluding remarks. Finally, an appendix that contains the proofs of the results exposed in Section 3 concludes the paper.

\setcounter{equation}{0}\setcounter{Def}{0}
\section{Semi-Markov chain with Borel state space}
\label{Pref}\setcounter{equation}{0} \setcounter{Def}{0}

The following is the minimal set of definitions covering discrete time semi-Markov chains with Borel state space necessary for the reading of the article. It should be remarked that very few studies deal with general Borel state space semi-Markov process and they concern mainly reliability models in continuous time, see e.g. Limnios and  Opri\c{s}an (2001), D'Amico (2011) and Limnios (2012).\\
\indent Let $(E, \mathcal{E})$ be a measurable space such that for all
$x\in E$ it results that $\{x\}\in \mathcal{E}$.
\begin{Def}[Sub-Markov Kernel]
A function $p(x,A)$, $x\in E$, $A\in \mathcal{E}$, is called a
sub-Markov kernel on $(E, \mathcal{E})$ if:\\
i) for every $x\in E$, $p(x,\cdot)$ is a measure on $\mathcal{E}$ such
that $p(x,E)\leq 1$;\\
ii) for every $A\in \mathcal{E}$, $p(\cdot, A)$ is a Borel measurable
function.
\end{Def}

\begin{Def}[Semi-Markov Kernel]
A function $Q(x,A,t)$, $x\in E$, $A\in \mathcal{E}$, $t\in \N$ is
called a discrete time semi-Markov kernel (SMK) on $(E, \mathcal{E})$
if:\\
i) $Q(x,A,\cdot)$ for all $x\in E$, $A\in \mathcal{E}$ is a
nondecreasing discrete real function such that $Q(x,A,0)=0$;\\
ii) $Q(\cdot, \cdot, t)$ for all $t\in \N$ is a sub-Markov kernel on
$(E, \mathcal{E})$;\\
iii) $p(\cdot, \cdot)=Q(\cdot, \cdot, \infty)$ is a Markov transition
probability function from $(E, \mathcal{E})$ to itself.
\end{Def}

For each $(x,s)\in E\times \N$, there exists a probability space
$(\Omega, \mathbb{F}, \mathbb{P}_{(x,s)})$ and a sequence of r.v.
$(J_{n}, T_{n})$ such that:
$$
\mathbb{P}_{(x,s)}[J_{0}=x, T_{0}=s]=1;
$$
\begin{equation*}
\begin{aligned}
& \mathbb{P}_{(x,s)}[J_{n+1}\in A, T_{n+1}\leq t \mid \mathbb{F}_{n}]\\& =\mathbb{P}_{(x,s)}[J_{n+1}\in A, T_{n+1}\leq t \mid J_{n},
T_{n}]=Q(J_{n}, A, t-T_{n}).
\end{aligned}
\end{equation*}
Thus $(J_{n}, T_{n})_{n\in \N}$ is a Markov process with state space
$E\times \N$ and transition probability function given by the
semi-Markov kernel $Q(x,A,t)$.

The sequence $(J_{n}, n\geq 0)$ gives the successive states of $E$ in
time and the sequence $(T_{n}, n\geq 0)$ gives the times at which
transitions occur.

Set $N(0)=0$ and define the process counting the number of transitions
that occur time $t$:
$$
N(t)=\sup\{n\in \N: T_{n}\leq t\},\,t>0.
$$

\begin{Def}[Semi-Markov chain]
The process $\{Z(t),\,t\in \N\}$, defined by 
$$
Z(t)=J_{N(t)}
$$
is a discrete time semi-Markov chain with arbitrary state space $E$
and kernel $Q(x,A,t)$.
\end{Def}
Let denote by 
$$
p(x,A):=\mathbb{P}[J_{n+1}\in A\mid J_{n}=x].
$$
It is simple to show that
$$
p(x,A)=\lim_{t\rightarrow \infty}Q(x,A,t).
$$
Define for any state $x\in E$ the probability
$$
H(x,t):=\mathbb{P}[T_{n+1}-T_{n}\leq t\mid J_{n}=x]=Q(x,E,t).
$$
It denotes the probability to leave state $x$ within time $t$ with a
transition toward any other state of $E$. Consequently, the quantity
$1-H(x,t)$ denotes the survival probability in state $x\in E$.

Inasmuch, $\forall x\in E$, $A\in \mathcal{E}$, $t\in \N$ $$
Q(x,A,t)\leq p(x,A),
$$ 
the measure $Q(x,\cdot , t)$ is absolutely continuous with respect to
$p(x,\cdot )$, then, according to the Radon-Nikodym theorem, it exists
a $\mathcal{E}$-measurable function $F(x,A,\cdot)$ such that
$$
Q(x,A,t)=\int_{A}F(x,y,t)p(x,dy).
$$
The function $F$ can be chosen as a distribution function expressing
the probability:
\begin{equation}
\label{effe}
F(x,y,t)=\mathbb{P}[T_{n+1}-T_{n}\leq t\mid J_{n}=x, J_{n+1}=y].
\end{equation}
\indent If the distribution function $(\ref{effe})$ is geometrically distributed, then $Z(t)$ is a Markov chain with measurable space $(E, \mathcal{E})$. In contrast, the semi-Markov chain environment allows the possibility to use any type of distribution function; as a result, semi-Markov chains supplies a much more general framework which is closer to the reality when compared with Markov chain based models.\\
\indent We denote by
\begin{equation}
q(x,A,t)=\mathbb{P}[J_{n+1}\in A, T_{n+1}-T_{n}=t \mid J_{n}=x],
\end{equation}
the probability to visit with next transition the set $A$ with a sojourn time length $t$ in the state $x$. Obviously it results that
\[
q(x,A,t)=\left\{
                \begin{array}{cl}
{Q(x,A,t)-Q(x,A,t-1)} & \mbox {if $t>0$,}\\
                       0   & \mbox {if $t= 0$.}
                \end{array}
          \right.
\] 
\begin{Def}[n-fold convolution of SMK]
Given the SMK $Q(x,A,t)$, define its n-fold convolution as follows:
\begin{equation}
\label{n-fold}
Q^{(n)}(x,A,t)=\left\{
                \begin{array}{cl}
{\int_{E}\sum_{s=1}^{t}Q(x,dy,s)Q^{(n-1)}(y,A,t-s)} & \mbox {if $n\geq
1, \, t> 0$,}\\
                       0   & \mbox {if $t\leq 0$,}
                \end{array}
          \right.
\end{equation}
where $Q^{(0)}(x,A,t)=\chi(x\in A)\chi(t>0)$ and $\chi(L)$ is the
indicator function of event $L$.
\end{Def}
It is simple to show that $\forall h\in \N$
$$
Q^{(n)}(x,A,t)=\mathbb{P}[J_{n+h}\in A, T_{n+h}-T_{h}\leq t\mid
J_{h}=x].
$$
The function 
$$
R(x,A,t):=\sum_{n=0}^{\infty}Q^{(n)}(x,A,t)=\mathbb{E}[N(t)\mid
J_{0}=x,T_{0}=0],
$$
is called Markov Renewal Function.
\begin{Def}[Normal SMK]
The SMK $Q(x,A,t)$ is said to be normal if $R(x,A,t)<\infty$ $\forall
x\in E$, $A\in \mathcal{E}$, $t\in \N$.
\end{Def}
It is important in semi-Markov modeling to introduce the following
auxiliary stochastic process:
\begin{Def}[Backward recurrence time process]
For each $t\in \N$ the process $\{B(t),\,t\in \N\}$ defined by 
$$
B(t)=t-T_{N(t)},
$$
is the backward recurrence time process
\end{Def}
The backward process denotes the time since the last transition and  is an important tool for detecting and quantifying the so called duration effect, namely the fact that the time the system is in a state influence its transition probabilities, see e.g. D'Amico et al. (2011).

\setcounter{equation}{0} \setcounter{Def}{0}
\section{The dividend valuation model}
\label{Pref} \setcounter{equation}{0} \setcounter{Def}{0}

Let denote by $P(k)$ the random variable denoting the fundamental price of the stock at time $k\in \N$, by $D(k)$ we denote the dividend at the same time and by $r$ one plus the required rate of return on the stock. We assume that $r$ is known and focus only on the other elements involved in the valuation of the stock.\\
\indent The fundamental equation affirms that
$p(k):=\mathbb{E}_{k}[P(k)]$ obey the following equation
\begin{equation}
\label{uno}
p(k)=\frac{\mathbb{E}_{(k)}[D(k+1)+P(k+1)]}{r},
\end{equation}
\noindent where $\mathbb{E}_{(k)}$ is the conditional expectation with
respect to the information available up to time $k$.\\
\indent As it is well known, see for example Samuelson \cite{samu73},
if we assume that
\begin{equation}
\label{trans1}
\lim_{i\rightarrow +\infty}\frac{\mathbb{E}_{(k)}[P(k+i)]}{r^{i}}=0,
\end{equation}
\noindent the unique solution of $(\ref{uno})$ is expressed by the series  
\begin{equation} 
\label{fund1}
p(k)=\sum_{i=1}^{+\infty}\frac{\mathbb{E}_{(k)}[D(k+i)]}{r^{i}}.
\end{equation}
\indent Formula $(\ref{fund1})$ valuates the stock as a function of
expected future dividend stream and discount rates.\\
\indent If condition $(\ref{trans1})$ is not assumed, Blanchard and
Watson \cite{blan82} proved that there can exist different solutions of
the fundamental equation revealing the presence of bubbles in the stock
market.\\
\indent The uncertainty in the dividend process propagates in the  price process.
To quantify this effect, it is possible to analyze the second order
moment of the price process. To this end we denote by
\begin{equation}
\label{secmom}
P^{2}(k)=\bigg(\frac{D(k+1)+P(k+1)}{r}\bigg)^{2}.
\end{equation}
\indent Successive substitutions for the future prices with future dividends in $(\ref{secmom})$ yield
\begin{equation}
\label{due}
\begin{aligned}
P^{2}(k)&=\sum_{i=1}^{N}\frac{D^{2}(k+i)}{r^{2i}}+2\sum_{i=1}^{N}\sum_{j>i}^{N}\frac{D(k+i)D(k+j)}{r^{i+j}}\\
&
+2\sum_{i=1}^{N}\frac{D(k+i)P(k+N)}{r^{i+N}}+\frac{P^{2}(k+N)}{r^{2N}}.
\end{aligned}
\end{equation}
\indent Therefore
\begin{equation}
\label{due}
\begin{aligned}
p^{2}(k):=\mathbb{E}[P^{2}(k)]&=\sum_{i=1}^{N}\frac{\mathbb{E}[D^{2}(k+i)]}{r^{2i}}+2\sum_{i=1}^{N}\sum_{j>i}^{N}\frac{\mathbb{E}[D(k+i)D(k+j)]}{r^{i+j}}\\
&
+2\sum_{i=1}^{N}\frac{\mathbb{E}[D(k+i)P(k+N)]}{r^{i+N}}+\frac{\mathbb{E}[P^{2}(k+N)]}{r^{2N}}.
\end{aligned}
\end{equation}
\indent Formula $(\ref{due})$ says that the risk (measured by the second order moment) of the price process is related to all future risks of the dividend process, to their product moments plus two additional terms that are related to future price and risk of the stock.\\
\indent If we assume that
\begin{equation}
\label{trans2a} 
\lim_{N\rightarrow
+\infty}\frac{\mathbb{E}_{(k)}[P^{2}(k+N)]}{r^{2N}}=0,
\end{equation}
and
\begin{equation}
\label{trans2b} 
\lim_{N\rightarrow
+\infty}\sum_{i=1}^{N}\frac{\mathbb{E}[D(k+i)P(k+N)]}{r^{i+N}}=0,
\end{equation}
then the unique solution of equation $(\ref{due})$ is given by
\begin{equation}
\label{fund2}
\mathbb{E}[P^{2}(k)]=\sum_{i=1}^{+\infty}\frac{\mathbb{E}[D^{2}(k+i)]}{r^{2i}}+2\sum_{i=1}^{+\infty}\sum_{j>i}\frac{\mathbb{E}[D(k+i)D(k+j)]}{r^{i+j}}.
\end{equation}
\indent We call expression $(\ref{fund2})$ the fundamental formula of the risk (second order moment of the price process). This formula expresses the risk as a function of the second order and product moments of the dividend process. The transversality conditions $(\ref{trans2a})$ and $(\ref{trans2b})$ eliminate the dependence of $p^{2}(k)$ on its own temporal path ($P^{2}(k+N)$). This means that conditions $(\ref{trans2a})$ and $(\ref{trans2b})$ are necessary in order to avoid rising in the risk ($p^{2}(k)$) due to an expected increase in the future risk ($P^{2}(k+N)$) and prices ($P(k+N)$) without any relation to the intrinsic risk of the dividend process. Therefore these conditions avoid the influence of future prices and risks on the present price and risk that are explained only in reference to the level of dividend and risk of the dividend process.

\subsection{The semi-Markov stock model}
\label{Pref}

As discussed in the introduction, a large literature deals with the
issue of evaluating formula $(\ref{fund1})$ under appropriate
assumptions about the dividend dynamic. Here we improve even further
this point advancing a more general model which imposes weaker
assumptions on the dividend process that allow effective computation of
the fundamental formulas $(\ref{fund1})$ and $(\ref{fund2})$ and
the fulfillment of the transversality conditions $(\ref{trans1})$,
$(\ref{trans2a})$ and $(\ref{trans2b})$.

Let us assume that the dividend process $\{D(k)\}_{k\in \N}$ obeys the
difference equation
\begin{equation}
\label{quattro}
D(k+1)=G(k+1)D(k),
\end{equation}
where the dividend growth factor $\{G(k)\}$ is described by a normal homogeneous semi-Markov chain with Borel state space $(E,\mathcal{E})$ and kernel $Q(x,A,t)$. It should be noted that model based on binomial, Markovian dynamics of the dividend are all embedded in the semi-Markov framework,
therefore all the results we will prove have particular cases that may be of practical interest.\\
\indent The model described by equation $(\ref{quattro})$ is discrete in time because dividends must be declared periodically
by the firm’s board of directors and therefore the problem has a natural description in term of discrete times. On the contrary the growth factor $\{G(k)\}$ can assume any real value and this is the reason for the choice of the general state space instead of the discrete one considered in D'Amico (2013).\\
\indent Inasmuch the semi-Markov process is time homogeneous, we can fix the current time $k=0$ without having any loss of generality. It is easy to verify that the process $(D(t),G(t),B(t))$ is a Markov process, then the fundamental equation becomes (see D'Amico (2013))
\begin{equation}
\label{cinque}
p(D(0),G(0),B(0))\!=\!\frac{\mathbb{E}_{(D(0),G(0),B(0))}[D(1)+P(D(1),G(1),B(1)]}{r}.
\end{equation}
\indent Therefore the current price depends on the current value of
dividends, on the value of the dividend growth process at that time and
on the duration in this state.\\
\indent Let us fix the initial condition at time $0$, $\{D(0)=d,G(0)=g,
B(0)=v\}$; the solution to equation $(\ref{cinque})$ under the
transversality condition
\begin{equation}
\label{sette}
\lim_{t\rightarrow
+\infty}\frac{\mathbb{E}_{(d,g,v)}[P(D(t),G(t),B(t)]}{r^{t}}=0,
\end{equation}
\noindent produces the results that the price equals the expected
present value of future dividends, i.e.
\begin{equation}
\label{sei}
p(d,g,v)=\sum_{t=1}^{+\infty}\frac{\mathbb{E}_{(d,g,v)}[D(t)]}{r^{t}}=\sum_{t=1}^{+\infty}\bigg(\frac{\mathbb{E}_{(d,g,v)}[\prod_{j=1}^{t}G(j)]}{r^{t}}\bigg)d.
\end{equation}
\indent Similarly, the fundamental formula of the second order moment
of the price process $(\ref{fund2})$ becomes:\begin{equation}
\begin{aligned}
\label{62}
&
p^{2}(d,g,v)=\sum_{i=1}^{+\infty}\bigg(\frac{\mathbb{E}_{(d,g,v)}[\prod_{j=1}^{t}G^{2}(j)]}{r^{2t}}\bigg)d^{2}\\
&
+2\sum_{t=1}^{+\infty}\sum_{w>t}\bigg(\frac{\mathbb{E}_{(d,g,v)}[\prod_{j=1}^{t}G^{2}(j)\prod_{j=t+1}^{w}G(j)]}{r^{t+w}}\bigg)d^{2},
\end{aligned}
\end{equation}
as long as the following conditions are satisfied:
\begin{eqnarray} 
\label{sette2}
\lim_{N\rightarrow
+\infty}\frac{\mathbb{E}_{(d,g,v)}[P^{2}(N)]}{r^{2N}}& = & 0,\\
\hspace{-0.8cm}\lim_{N\rightarrow
+\infty}\sum_{i=1}^{N}\frac{\mathbb{E}_{(d,g,v)}[D(i)P(N)]}{r^{i+N}}& =
& 0.
\end{eqnarray}
\indent The computation and the convergence of formulas $(\ref{sei})$
and $(\ref{62})$ needs the study of the (product) growth dividend
process.
\begin{Def}[The growth dividend process]
\label{DEF3.1}
The process $\big\{A_{g,v}^{(k)}(t),\,t\in \N, g\in E, v\in \N, k\in \N
\big\}$, defined for $t>0$ by
\begin{equation}
A_{g,v}^{(k)}(t;\omega)=\left\{
                \begin{array}{cl}
{\prod_{j=1}^{t}G^{k}(j;\omega)} & \mbox {if $\omega \in
\Omega_{g,v}$,}\\
                       0   & \mbox {otherwise,}
                \end{array}
          \right.
\end{equation}
and for $t=0$ by
\begin{equation}
A_{g,v}^{(k)}(0;\omega)=\left\{
                \begin{array}{cl}
                       {1} & \mbox {if $\omega \in \Omega_{g,v}$,}\\
                       0   & \mbox {otherwise,}
                \end{array}
          \right.
\end{equation}
where 
\begin{equation*}
\Omega_{g,v}=\{\omega \in \Omega: G(0,\omega)=g, B(0;\omega)=v\},
\end{equation*}
 is called the growth dividend process.
\end{Def} 
Thus, $\forall t>0$ and $\forall a\in \R$ the random variable
$A_{g,v}^{(k)}(t)$ has the following distribution function
\begin{equation}
\mathbb{P}[A_{g,v}^{(k)}(t)\leq
a]=\mathbb{P}[\prod_{j=1}^{t}G^{k}(j)\leq a\mid G(0)=g,B(0)=v].
\end{equation}
\begin{Def}[The product growth dividend process]
The process $\big\{AP_{g,v}^{(k,w)}(t,s),\, s>t, g\in E, v\in
\N,\,k,w\in \N \big\}$, defined for $t>0$ by
\begin{equation}
\label{AP}
AP_{g,v}^{(k,w)}(t,s)=A_{g,v}^{(k)}(t)\cdot A_{G(t),B(t)}^{(w)}(s)
\end{equation}
 is called the product growth dividend process.
\end{Def} 
Thus, $\forall s>t>0$ and $\forall a\in \R$ the random variable
$AP_{g,v}^{(k,w)}(t,s)$ has the following distribution function
\begin{equation}
\mathbb{P}[AP_{g,v}^{(k,w)}(t,s)\leq
a]=\mathbb{P}[\prod_{j=1}^{t}G^{k}(j)\prod_{j=t+1}^{s}G^{w}(j)\leq
a\mid G(0)=g,B(0)=v].
\end{equation}
We denote the corresponding expectations by
\begin{eqnarray} 
M_{g,v}^{(k)}(t)& = &\mathbb{E}_{(d,g,v)}[A_{g,v}^{(k)}(t)],
 \nonumber\\
\hspace{-0.8cm}M_{g,v}^{(k,w)}(t,s)& =& \mathbb{E}_{(d,g,v)}[A_{g,v}^{(k)}(t)A_{G(t),B(t)}^{(w)}(s)]. \nonumber
\end{eqnarray}
\begin{Proposition}[Moments of the product growth dividend process]
\label{mom2}
For all $g\in E$, $v,t,s\in \N$, the product moment
$M_{g,v}^{(k,w)}(t,s)$ satisfies the following equation:
\begin{equation}
\label{product moment}
\begin{aligned}
& M_{g,v}^{(k,w)}(t,s)=\bigg(\frac{1-H(g,v+s)}{1-H(g,v)}\bigg)
(g)^{ws+t(k-w)}\\
& +\sum_{\theta=t+1}^{s}\int_{E}\frac{\dot{q}(g,y, \theta+v)}{1-H(g,v)}
(g)^{w\theta+t(k-w)-w}\,y^{w}\, M_{y,0}^{(w)}(s-\theta)dy\\
& +\sum_{\theta=1}^{t}\int_{E}\frac{\dot{q}(g,y, \theta+v)}{1-H(g,v)}
(g)^{k(\theta-1)}\,y^{k}\, M_{y,0}^{(k,w)}(t-\theta,s-\theta)dy.
\end{aligned}
\end{equation}
\end{Proposition}
\begin{Proof}
See the appendix.
\end{Proof}
\begin{Corollary}[Moments of the growth dividend process]
\label{mom1}
For all $g\in E$, $v,t\in \N$, the k-order moment of the dividend
growth-product process satisfies the following equation:
\begin{equation}
\begin{aligned}
\label{Iorder}
M_{g,v}^{(k)}(t)&=\bigg(\frac{1-H(g,v+t)}{1-H(g,v)}\bigg) (g)^{kt}\!\\
& +\sum_{\theta=1}^{t}\int_{E}\frac{\dot{q}(g,y, \theta+v)}{1-H(g,v)}
(g)^{k(\theta-1)}\,y^{k}\, M_{y,0}^{(k)}(t-\theta)dy.\\
& M_{g,v}^{(k)}(0)=1.
\end{aligned}
\end{equation}
\end{Corollary}
\begin{Proof}
See the appendix.
\end{Proof}\\
\indent Equations $(\ref{product moment})$ and $(\ref{Iorder})$ are of recursive type and therefore can recursively be solved as discussed later in the subsection on computational methods. The only unknown parameters of these equations are $M_{g,v}^{(k,w)}(t,s)$ and $M_{g,v}^{(k)}(t)$.\\
\indent Equation $(\ref{product moment})$ is divided in three different parts, the first corresponds to the event that no transition in the growth dividend process is made, the second the event that next transition is made at $\theta \in \{t+1,\ldots , s\}$, the third term includes the event that next transition of the growth dividend process is made at  $\theta \in \{1,\ldots , t\}$.\\
\indent Moreover notice that the solution of equation $(\ref{product moment})$ necessitates before the resolution of equation $(\ref{Iorder})$ because the second term of $(\ref{product moment})$ contains the factor $M_{g,v}^{(w)}(s-\theta)$ which is the unknown of equation $(\ref{Iorder})$.\\
\indent Proposition $\ref{mom2}$ and Corollary $\ref{mom1}$ provide the following representations of prices and risks:
\begin{equation}
\label{nuova}
p(d,g,v)=\sum_{t=1}^{+\infty}\frac{M_{g,v}^{(1)}(t)}{r^{t}}d.
\end{equation}
\begin{equation}
\label{nuova2}
p^{2}(d,g,v)=\sum_{t=1}^{+\infty}\frac{M_{g,v}^{(2)}(t)}{r^{2t}}d^{2}+2\sum_{t=1}^{+\infty}\sum_{s>t}\frac{M_{g,v}^{(2,1)}(t,s)}{r^{t+s}}d^{2}.
\end{equation}
Major questions are about the convergence of these series and on
computational methods. 

\subsection{Finiteness of prices and risks of the semi-Markov stock model}
\label{convergence} 

Let us consider now the question of the convergence of the series $(\ref{nuova})$ and $(\ref{nuova2})$. To this purpose, let denote by
\begin{equation*}
\begin{aligned}
\label{settebis}
&\overline{g}(v)=\sup_{y\in
E}\Bigg(y\frac{1-H(y,v+1)}{1-H(y,v)}+\int_{E}x
\frac{\dot{q}(y,x,v+1)}{1-H(y,v)}dx\Bigg),
\end{aligned}
\end{equation*}
\noindent and by 
\begin{equation*}
\label{condition}
\overline{g}=\sup_{v\in \N}\overline{g}(v).
\end{equation*}
with the convention $\frac{0}{0}=0$.\\
\indent All the work in the rest of the article will be done under the assumption:
\begin{equation}
\label{A1}
\hspace{-7.0cm}A1: \hspace{5.8cm}\overline{g}<r.
\end{equation}

\begin{Theorem}(Finiteness of prices)
\label{teo1}
Under assumption A1 it results that:\\
i) the series
$p(d,g,v)=\sum_{t=1}^{+\infty}\frac{\mathbb{E}_{(d,g,v)}[D(t)]}{r^{t}}$
converges\\
ii) it meets the asymptotic condition 
\begin{equation}
\label{asymp}
\lim_{t\rightarrow +\infty}\frac{
\mathbb{E}_{(d,g,v)}[P(D(t),G(t),B(t)]}{r^{t}}=0.
\end{equation}
\end{Theorem}
\begin{Proof}
See the appendix.
\end{Proof}\\
\indent This theorem is the generalization of Theorem 2 in D'Amico (2013) where the dividend process was described by a finite state space semi-Markov chain. The result can be further generalized to cover the second order moment of the price process if an additional assumption, namely A2, is formulated:
\begin{equation}
\label{A2}
\hspace{-6.5cm}A2: \hspace{5.8cm}\overline{g}^{(2)}<r^{2},
\end{equation}
where $\overline{g}^{(2)}=\sup_{v\in \N}\overline{g}^{(2)}(v)$ and
$\overline{g}^{(2)}(v)=\sup_{y\in
E}\Bigg(y^{2}\frac{1-H(y,v+1)}{1-H(y,v)}+\int_{E}x^{2}
\frac{\dot{q}(y,x,v+1)}{1-H(y,v)}dx\Bigg)$.
\begin{Theorem}(Finiteness of risks)
\label{theo2}
Under assumptions A1 and A2 the series
\begin{equation*}
\label{sei2}
p^{2}(d,g,v)=\sum_{i=1}^{+\infty}\frac{\mathbb{E}_{(d,g,v)}[D^{2}(i)]}{r^{2i}}+2\sum_{i=1}^{+\infty}\sum_{j>i}\frac{\mathbb{E}_{(d,g,v)}[D(i)D(j)]}{r^{i+j}},
\end{equation*}
converges and meets the asymptotic conditions: 
\begin{equation} 
\label{sette2}
\lim_{N\rightarrow
+\infty}\frac{\mathbb{E}_{(d,g,v)}[P^{2}(N)]}{r^{2N}} = 0,
\end{equation}
\begin{equation}
\label{sette3}
\lim_{N\rightarrow
+\infty}\sum_{i=1}^{N}\frac{\mathbb{E}_{(d,g,v)}[D(i)P(N)]}{r^{i+N}}
= 0.
\end{equation}
\end{Theorem}
\begin{Proof}
See the appendix.
\end{Proof}\\
\indent These two theorems present the assumptions under which the
transversality conditions $(\ref{asymp})$, $(\ref{sette2})$,
$(\ref{sette3})$ are satisfied. This avoids the presence of speculative
bubbles and therefore permits the representation of prices and risks as series that are shown to be convergent and that depend only on the fundamental variables (the dividend process).\\
\indent Assumption A2 is necessary in determining the finiteness of risks because it controls the second order moment of the growth dividend process. The only assumption A1 is unable to guarantee the finiteness of both prices and risks because it does not imply the A2. To fix this point we can consider a clarifying example where the dividend growth process obeys a two state discrete time Markov chain with transition probability matrix
\begin{displaymath}
{\mathbf{P}}=
{\left[
\begin{matrix}
0.6  & 0.4\\
0.2 & 0.8
\end{matrix}
\right],}
\end{displaymath}
and state space $E=\{1, 1.5\}$. Set $\mathbf{g}=[1, 1.5]^{'}$, $r=1.41$ and compute $\mathbf{P}\cdot \mathbf{g}=[1.2, 1.4]^{'}$. Then $\overline{g}=1.4$ and the assumption A1 is satisfied. Anyway $\mathbf{g}^{2}=[1, 2.25]^{'}$ and $\mathbf{P}\cdot \mathbf{g}^{2}=[1.5, 2]^{'}$ which implies $\overline{g}^{(2)}=2$. A simple comparison shows that although A1 is satisfied, A2 is not.

\subsection{Computational methods}

In this subsection we propose two methods to compute the fundamental price and the fundamental risk as represented by formulas $(\ref{nuova})$ and $(\ref{nuova2})$.\\
\indent The first method is a direct computation of $(\ref{nuova})$ and $(\ref{nuova2})$ based on Proposition $(\ref{mom2})$ and Corollary $(\ref{mom1})$ based on numerical approximation of the considered integral equations. This method was originally considered by Janssen and Manca (2004) for computing the transition probability function of a continuous time finite state space semi-Markov process. Here we modify it to be useful in evaluating the moments of the (product) growth dividend process.\\
Let us consider before the equation $(\ref{Iorder})$. As already remarked, this is a recursive equation which can be solved as explained here below.\\
\indent Consider a state space grid with discretization step $h$:
\begin{equation}
\label{grid}
\omega =\{x(i)=ih\}
\end{equation}
where $i=d,d+1,\ldots,N$, $dh=\inf E$, $Nh=\sup E$.\\
\indent Now consider a generic quadrature formula to approximate the integral on the grid $\omega$ to be applied $\forall v\in\N$, $\forall t\in \N$:
\begin{equation*}
\begin{aligned}
& \int_{E}\frac{\dot{q}(g,y, \theta+v)}{1-H(g,v)}
(g)^{k(\theta-1)}\,y^{k}\, M_{y,0}^{(k)}(t-\theta)dy\\
& \approx \sum_{l=d}^{N}w_{dN}(l)\,(ah)^{k(\theta-1)}\,(lh)^{k} M_{lh,0}^{(k)}(t-\theta)\, \frac{\dot{q}(ah,lh, \theta+v)}{1-H(ah,v)},
\end{aligned}
\end{equation*}
where $ah$ is the approximate value of $g$ on the grid $(\ref{grid})$,  $\dot{q}(ah,lh, \theta+v)$ is the discrete derivative of $q$ and $w_{dN}(\cdot)$ are the weights relative to the quadrature formula.\\
\indent Therefore by substitution we can approximate the equation $(\ref{Iorder})$ with the discrete equation:
\begin{equation}
\label{discreto}
\begin{aligned}
&  M_{ah,v}^{(k)}(t)=\frac{1-H(ah,v+t)}{1-H(ah,v)}(ah)^{kt}\\
& + \sum_{\theta =1}^{t}\sum_{l=d}^{N}w_{dN}(l)\,(ah)^{k(\theta-1)}\,(lh)^{k} M_{lh,0}^{(k)}(t-\theta)\, \frac{\dot{q}(ah,lh, \theta+v)}{1-H(ah,v)}.
\end{aligned}
\end{equation}
\indent The top system of equations can also be written more compact using matrix notation. Let ${\mathbf{M}}_{v}^{(k)}(t)$ be the $|\omega|\times 1$ column vector of elements
\begin{equation}
 {\mathbf{M}}_{v}^{(k)}(t)=[M_{dh,v}^{(k)}(t), M_{(d+1)h,v}^{(k)}(t),\ldots , M_{Nh,v}^{(k)}(t)]^{t},
\end{equation}
which stores the discretized values of the $k$-order moment of the growth dividend process. The quantity $|\omega|$ denotes the cardinality of the grid $(\ref{grid})$.\\
\indent Let ${\mathbf{D}}_{v}(t)$ be the diagonal matrix of dimension $|\omega|\times |\omega|$ with elements:
\begin{displaymath}
{\mathbf{D}}_{v}(t)=
{\left[
\begin{matrix}
\frac{1-H(dh,v+t)}{1-H(dh,v)}  & 0 & \dots & 0\\
0 & \frac{1-H((d+1)h,v+t)}{1-H((d+1)h,v)}  & \dots & 0\\
\vdots & \vdots  & \ddots & \vdots\\
0 & 0 & \dots & \frac{1-H(Nh,v+t)}{1-H(Nh,v)}
\end{matrix}
\right]}.
\end{displaymath}
\indent Define by $\mathbf{A}^{(k)}(t)$ the diagonal matrix with elements:
\begin{displaymath}
{\mathbf{A}}^{(k)}(t)=
{\left[
\begin{matrix}
(dh)^{kt}  & 0 & \dots & 0\\
0 & ((d+1)h)^{kt}  & \dots & 0\\
\vdots & \vdots  & \ddots & \vdots\\
0 & 0 & \dots & (Nh)^{kt}
\end{matrix}
\right]},
\end{displaymath}
and by $\mathbf{B}_{v}(t)$ the $|\omega|\times |\omega|$ matrix of generic element $\mathbf{B}_{v}(t)=\Big(w_{dN}(l)\frac{\dot{q}(ah,lh,t+v)}{1-H(ah,v)}\Big)_{ah,lh\in \omega}$.\\
\indent Then the recursive equations $(\ref{discreto})$ can be written in our new matrix-notation as,
\begin{equation}
\label{matrix}
 \begin{aligned}
\mathbf{M}_{v}^{(k)}(t)&=\mathbf{D}_{v}(t)\cdot \mathbf{A}^{(k)}(t)\cdot \mathbf{1}_{|\omega|}\\
& +\sum_{\theta =1}^{t}\big(\mathbf{B}_{v}(\theta)\lozenge \mathbf{A}^{(k)}(t)\big)\cdot \big((\mathbf{A}^{(k)}(1)\cdot \mathbf{1}_{|\omega|})\lozenge \mathbf{M}_{0}^{(k)}(t-\theta)\big),
\end{aligned}
\end{equation}
where $\mathbf{1}_{|\omega|}$ is the $|\omega|\times 1$ vector with all components equal $1$, the symbol $\cdot$ is the ordinary row by column matrix product and $\lozenge$ is the Hadamard or element-wise multiplication of matrices.\\
\indent The  solution of $(\ref{matrix})$ is straightforward because for $t=0$ we know from Corollary $(\ref{mom1})$ that $\mathbf{M}_{v}^{(k)}(0)=\mathbf{1}_{|\omega|}$ $\forall v\in \N$ and $\forall k\in \N$. Then set $t=1$ to have $\forall v\in \N$
\begin{equation}
\label{matrix2}
\mathbf{M}_{v}^{(k)}(1)=\mathbf{D}_{v}(1)\cdot \mathbf{A}^{(k)}(1)\cdot \mathbf{1}_{|\omega|}
+\big(\mathbf{B}_{v}(1)\lozenge \mathbf{A}^{(k)}(1)\big)\cdot \big(\mathbf{A}^{(k)}(1)\cdot \mathbf{1}_{|\omega|}\big).
\end{equation}
\indent All the terms on the right hand side are known and then it is possible to compute $\mathbf{M}_{v}^{(k)}(1)$. Once we know $\mathbf{M}_{v}^{(k)}(1)$ we can proceed to compute $\mathbf{M}_{v}^{(k)}(2)$ and in general knowing $\mathbf{M}_{v}^{(k)}(0)$, $\mathbf{M}_{v}^{(k)}(1)$,...,$\mathbf{M}_{v}^{(k)}(t-1)$ it is possible to compute $\mathbf{M}_{v}^{(k)}(t)$.\\
\indent Similar arguments can be used to solve equation $(\ref{product moment})$. Once the equations $(\ref{Iorder})$ and $(\ref{product moment})$ are solved it is possible to evaluate the prices and the risks through formulas $(\ref{nuova})$ and  $(\ref{nuova2})$, respectively. Anyway here below we do not report all the necessary computation because we propose an alternative procedure for computing prices and risks since 
the direct computation is not the most convenient way to proceed. The above valuation formulas $(\ref{nuova})$ and $(\ref{nuova2})$ can be conveniently represented by introducing two auxiliary functions.
\begin{Def}[price-dividend ratio]
For all $g\in E$, $v\in \N$ the price-dividend ratio is the function
defined by:
\begin{equation}
\psi_{1}(v, g)=\sum_{t=1}^{+\infty}\frac{M_{g,v}^{(1)}(t)}{r^{t}}.
\end{equation}
\end{Def}
Hence, according to equation $(\ref{nuova})$, the former definition
provides a compact representation of the price:
\begin{equation}
\label{danumerare}
p(d,g,v)=\psi_{1}(v, g)d.
\end{equation}
\begin{Def}[second order price-dividend ratio]
For all $g\in E$, $v\in \N$ the second order price-dividend ratio is
the function defined by:
\begin{equation}
\psi_{2}(v,
g)=\Bigg(\sum_{t=1}^{+\infty}\frac{M_{g,v}^{(2)}(t)}{r^{2t}}+2\sum_{t=1}^{+\infty}\sum_{s>t}\frac{M_{g,v}^{(2,1)}(t,s)}{r^{t+s}}\Bigg).
\end{equation}
\end{Def}
Hence, according to equation $(\ref{nuova2})$, the former definition provides a compact representation of the risk:
\begin{equation}
\label{psi2}
p^{2}(d,g,v)=\psi_{2}(v, g)d^{2}.
\end{equation}
\begin{Proposition}[price-dividend ratio equation]
\label{prop3}
Under assumption A1 the price dividend ratio satisfies for all $g\in E$ and $v\in \N$ the following system of equations:
\begin{equation}
\label{sistema}
\begin{aligned}
& \frac{1-H(g,v+1)}{1-H(g,v)} g \psi_{1}(v+1, g)-r\psi_{1}(v, g)\\
& + \int_{E}\psi_{1}(0, x)x\frac{\dot{q}(g,x,v+1)}{1-H(g, v)}dx = -
E(g,v).
\end{aligned}
\end{equation} 
where
$E(g,v)=\mathbb{E}_{(d,g,v)}[G(1)]$.
\end{Proposition}
\begin{Proof}
See the appendix.
\end{Proof}
\begin{Proposition}[second order price-dividend ratio equation]
\label{prop4}
Under assumptions A1 and A2 the second order price-dividend ratio satisfies for all $g\in E$ and $v\in \N$ the
following system of equations:
\begin{equation}
\label{sistema2}
\begin{aligned}
& \frac{1-H(g,v+1)}{1-H(g,v)} g^{2} \psi_{2}(v+1, g)-r^{2}\psi_{2}(v,
g)\\
& + \int_{E}{\psi_{2}(0, x)}x^{2}\frac{\dot{q}(g,x,v+1)}{1-H(g, v)}dx +
\frac{1-H(g,v+1)}{1-H(g,v)} g^{2}{\psi_{1}(v+1, g)}\\
& + \int_{E}{\psi_{1}(0, x)}x^{2}\frac{\dot{q}(g,x,v+1)}{1-H(g, v)}dx
 = - E^{2}(g,v).
\end{aligned}
\end{equation} 
where
$E^{2}(g,v)=\mathbb{E}_{(d,g,v)}[G^{2}(1)]$.
\end{Proposition}
\begin{Proof}
See the appendix.
\end{Proof}\\
\indent The above propositions can be used to compute the fundamental price and the fundamental risk. We describe the methodology only for the price because similar considerations hold also for the risk.\\
\indent As a first step it is necessary to evaluate $\psi_{1}(0,
x)=\sum_{t=1}^{+\infty}\frac{M_{x,0}^{(1)}(t)}{r^{t}}$, this can be done by using the truncated series development. This is possible because we proved that
the series $\sum_{t=1}^{+\infty}\frac{M_{x,0}^{(1)}(t)}{r^{t}}$ is
convergent and then in principle we can control the error of
approximation. Therefore, fixed an error $\epsilon>0$ we can find an
integer $T$ such that
\[
\bigg|\sum_{t=1}^{+\infty}\frac{M_{x,0}^{(1)}(t)}{r^{t}}-\sum_{t=1}^{T}\frac{M_{x,0}^{(1)}(t)}{r^{t}}\bigg|<\epsilon
.
\]
\indent The computation of $\sum_{t=1}^{T}\frac{M_{x,0}^{(1)}(t)}{r^{t}}$ requires the solution of $(\ref{matrix})$ only for $v=0$.\\
\indent As a second step we approximate the integral
$\int_{E}\psi_{1}(0, x)x\frac{\dot{q}(g,x,v+1)}{1-H(g, v)}dx$ inside
equation $(\ref{sistema})$ by replacing to $\psi_{1}(0,x)$ its truncation approximation. This can be accomplished as follows:
\begin{equation}
\begin{aligned}
\label{integral}
\int_{E}\psi_{1}(0, x)x\frac{\dot{q}(g,x,v+1)}{1-H(g, v)}dx & \approx
\sum_{t=1}^{T}\int_{E}\frac{M_{x,0}^{(1)}(t)}{r^{t}}x\frac{\dot{q}(g,x,v+1)}{1-H(g,
v)}dx\\
& \approx
\sum_{t=1}^{T}\sum_{l=0}^{k}w(l)\frac{M_{lh,0}^{(1)}(t)}{r^{t}}\cdot
lh\cdot \frac{\dot{q}(g,lh,v+1)}{1-H(g, v)}=:K(g, v),
\end{aligned}
\end{equation}
\noindent where the last approximation is obtained by using a general
quadrature formula after discretization of the phase space $E$ with a
discretization step length $h$. It is important to note that the quadrature formula is applied here to evaluate a quantity $K(g,v)$ that is independent of $t$. On the contrary, the direct method requires the computation of the quadrature formula for all times $t$.\\
\indent The third step consists in substituting the quantity $K(g,v)$
inside equation $(\ref{sistema})$ and to rearrange terms to obtain the
following first order difference equation:
\begin{equation}
\label{diffeq}
\psi_{1}(v+1, g) = z(v, g,r)\psi_{1}(v, g)+\gamma(v, g,r),
\end{equation}
where 
\begin{eqnarray*}
z(v, g,r) & = & \Big[ \frac{r(1-H(g,v))}{g(1-H(g,v+1))}\Big],\\
\gamma(v, g,r) & = &\frac{(K(g,v)+E(g,v))(1-H(g,v))}{g(1-H(g,v+1))}.
\end{eqnarray*}
\indent The solution of the inhomogeneous first order linear difference
equation $(\ref{diffeq})$ is given by
\begin{equation*}
\label{soldiffeq}
\psi_{1}(v+1, g) = \Big( \prod_{j=0}^{v} z(j, g,r) \Big) \Big(
\psi_{1}(0, g)+ \sum_{k=0}^{v} \frac{\gamma(k, g,r)}{\prod_{j=0}^{v}
z(j, g,r)} \Big).
\end{equation*}
\indent Formula $(\ref{danumerare})$ provides the price-dividend ratio
for all possible states $g$ and backward values $v$ and therefore gives
the possibility to recover prices simply through a multiplication with
the value of the dividend process:
$$
p(d,g,v)=\Big( \prod_{j=0}^{v} z(j, g,r) \Big) \Big( \psi_{1}(0, g)+
\sum_{k=0}^{v} \frac{\gamma(k, g,r)}{\prod_{j=0}^{v} z(j, g,r)} \Big)d.
$$

\setcounter{equation}{0}\setcounter{Def}{0}
\section{Conclusions}
\label{Pref}\setcounter{equation}{0} \setcounter{Def}{0}

Many studies deal with pricing firms on the basis of fundamentals. This paper shows how to compute prices and risks when the dividend dynamic is driven by a discrete time semi-Markov chain with general state space. Transversality conditions that avoids bubbles in the market and that guarantees the finiteness of price and risk are established. Moreover different computational methods are supplied for the implementation of the model.\\
\indent The proposed model encompasses the majority of the discounted dividend valuation models and therefore furnishes very general results in this field.\\
\indent Possible avenues for future developments of our model include:

a) the application of the model to real dividend data which necessitates the developing of techniques for the estimation of the model parameters;

b) Estimation of the required rate of return by adopting an appropriate stochastic model.

\setcounter{equation}{0}\setcounter{Def}{0}
\section*{Appendix: proofs}
\label{Pref}\setcounter{equation}{0} \setcounter{Def}{0}

\noindent {\bf{Proof} of Proposition $(\ref{mom2})$
}\\
Let us assume, without loss of generality, that $N(0)=0$. Then, because
the event $\{T_{1}>s\}$, $\{T_{1}\in (t,s]\}$ and $\{T_{1}\leq t\}$ are
disjoint it results that:
\begin{equation}
\label{M}
\begin{aligned}
M_{g,v}^{(k,w)}(t,s)&=\mathbb{E}[AP_{g,v}^{(k,w)}(t,s)1_{\{T_{N(0)+1}>s\}}]+\mathbb{E}[AP_{g,v}^{(k,w)}(t,s)1_{\{t<T_{N(0)+1}\leq
s\}}]\\
& +\mathbb{E}[AP_{g,v}^{(k,w)}(t,s)1_{\{T_{N(0)+1}\leq t\}}].
\end{aligned}
\end{equation}
Let us evaluate the three expectations above.\\
\indent In the case $T_{N(0)+1}>s$ there will be any transition up to
time $s$ and therefore relation $(\ref{AP})$ modifies as follows:
\begin{equation}
AP_{g,v}^{(k,w)}(t,s)=A_{g,v}^{(k)}(t)\cdot A_{G(t),B(t)}^{(w)}(s)= \prod_{i=1}^{t}g^{k}\cdot \prod_{j=t+1}^{s}g^{w}=g^{ws+t(k-w)}.
\end{equation}
\indent The event $T_{N(0)+1}>s$ occurs with probability
\begin{equation}
\label{prob1}
\begin{aligned}
& \mathbb{P}[T_{N(0)+1}>s \mid J_{N(0)}=g, T_{N(0)}=-v, T_{N(0)+1}>0]\\& =\frac{\mathbb{P}[T_{N(0)+1}>s \mid J_{N(0)}=g,
T_{N(0)}=-v]}{\mathbb{P}[T_{N(0)+1}>0 \mid J_{N(0)}=g, T_{N(0)}=-v]}
=\frac{1-H(g,s+v)}{1-H(g,v)}.
\end{aligned}
\end{equation}
\indent Then it results that
\begin{equation}
\begin{aligned}
\label{primo}
\mathbb{E}[AP_{g,v}^{(k,w)}(t,s)1_{\{T_{N(0)+1}>s\}}]&=\mathbb{E}[A_{g,v}^{(k)}(t)A_{G(t),B(t)}^{(w)}(s)1_{\{T_{N(0)+1}>s\}}]\\
&=\frac{1-H(g,s+v)}{1-H(g,v)}g^{ws+t(k-w)}.
\end{aligned}
\end{equation}
\indent In the case when $T_{N(0)+1}\in (t,s]$ since 
$AP_{g,v}^{(k,w)}(t,s)=A_{g,v}^{(k)}(t)\cdot A_{G(t),B(t)}^{(w)}(s),
$ and because any transition occurs until time $t$, we have
\begin{equation}
\label{alfa}
A_{g,v}^{(k)}(t)= \prod_{j=1}^{t}G^{k}(j)=\prod_{j=1}^{t}g^{k}=g^{kt}.
\end{equation}
To evaluate $A_{G(t),B(t)}^{(w)}(s)$ consider also the state occupied
with next transition, say $J_{N(0)+1}$. Thus
\begin{equation}
\label{beta}
\begin{aligned}
A_{G(t),B(t)}^{(w)}(s)&=\prod_{j=t+1}^{s}G^{w}(j)=\prod_{j=t+1}^{T_{N(0)+1}-1}G^{w}(j)\cdot
J_{N(0)+1}^{w}\cdot A_{J_{N(0)+1},0}^{(w)}(s-T_{N(0)+1})\\
& =g^{w(T_{N(0)+1}-t-1)}\cdot J_{N(0)+1}^{w}\cdot
A_{J_{N(0)+1},0}^{(w)}(s-T_{N(0)+1}).
\end{aligned}
\end{equation}
Multiply $(\ref{alfa})$ and $(\ref{beta})$ to obtain:
\begin{equation}
\label{gamma}
AP_{g,v}^{(k,w)}(t,s)=g^{wT_{N(0)+1}+t(k-w)-w}\cdot J_{N(0)+1}^{w}
\cdot A_{J_{N(0)+1},0}^{(w)}(s-T_{N(0)+1}).
\end{equation}
The expectation of the top process can be computed once the probability
of the conditional event
\[
\{J_{N(0)+1}=y,T_{N(0)+1}=\theta \in \{t+1,\ldots ,s\} \mid J_{N(0)}=g,
T_{N(0)}=-v, T_{N(0)}>0\}
\] 
is known. Thus,
\begin{equation}
\label{prob2}
\begin{aligned}
& \mathbb{P}[J_{N(0)+1}\in (y,y+dy), T_{N(0)+1}=\theta \in \{t+1,\ldots
, s\} \mid J_{N(0)}=g, T_{N(0)}=-v, T_{N(0)+1}>0]\\
& =\frac{\mathbb{P}[J_{N(0)+1}\in (y,y+dy), T_{N(0)+1}=\theta,
T_{N(0)+1}>0 \mid J_{N(0)}=g, T_{N(0)}=-v]}{\mathbb{P}[T_{N(0)+1}>0,
T_{N(0)}=-v \mid J_{N(0)}=g]}\\
& =\frac{\mathbb{P}[J_{N(0)+1}\in (y,y+dy), T_{N(0)+1}-T_{N(0)}=\theta
+v \mid J_{N(0)}=g]}{\mathbb{P}[T_{N(0)+1}-T_{N(0)}>v \mid
J_{N(0)}=g]}\\
& =\frac{\dot{q}(g,y,\theta +v)dy}{1-H(g,v)}.
\end{aligned}
\end{equation}
Notice that the random variable $A_{J_{N(0)+1},0}^{(w)}(s-T_{N(0)+1})$
is independent of the joint random variable $(J_{N(0)+1},T_{N(0)+1})$
because the product process $A_{J_{N(0)+1},0}^{(w)}(s-T_{N(0)+1})$ has
the Markov property at transition times and consequently once the state
$J_{N(0)+1}$ and the time $T_{N(0)+1}$ are known, its behavior does not
depends on the distribution of $(J_{N(0)+1},T_{N(0)+1})$. Then, by
taking the expectation of $(\ref{gamma})$ we get:
\begin{equation}
\label{secondo}
\mathbb{E}[AP_{g,v}^{(k,w)}(t,s)1_{\{t<T_{N(0)+1}\leq
s\}}]=\sum_{\theta =t+1}^{s}\int_{E}\frac{\dot{q}(g,y,\theta
+v)}{1-H(g,v)}\cdot g^{w\theta+t(k-w)-w}\cdot y^{w} \cdot
M_{y,0}^{(w)}(s-\theta) dy.
\end{equation}
\indent The final case is when $T_{N(0)+1}<t$. The product growth
dividend process becomes:
\begin{equation}
\begin{aligned}
\label{delta}
&AP_{g,v}^{(k,w)}(t,s)=A_{g,v}^{(k)}(t)\cdot A_{G(t),B(t)}^{(w)}(s)\\
& =\prod_{j=1}^{T_{N(0)+1}-1}g^{k}\cdot J_{N(0)+1}^{k} \cdot
A_{J_{N(0)+1},0}^{(k)}(t-T_{N(0)+1}) \cdot
A_{G(t-T_{N(0)+1}),B(t-T_{N(0)+1})}^{(w)}(s-T_{N(0)+1})\\
& =g^{k(T_{N(0)+1}-1)}\cdot J_{N(0)+1}^{k} \cdot
AP_{J_{N(0)+1},0}^{(k,w)}(t-T_{N(0)+1},s-T_{N(0)+1}).
\end{aligned}
\end{equation}
\indent The expectation of $(\ref{delta})$ can be computed considering
the probability of the conditional event
\[
\{J_{N(0)+1}\in (y, y+dy) ,T_{N(0)+1}=\theta \leq t \mid J_{N(0)}=g,
T_{N(0)}=-v, T_{N(0)}>0\}
\] 
that coincides with $\frac{\dot{q}(g,y,\theta +v)dy}{1-H(g,v)}$ and the
fact that the random variable
$AP_{J_{N(0)+1},0}^{(k,w)}(t-T_{N(0)+1},s-T_{N(0)+1})$ is independent
of the joint random variable $(J_{N(0)+1},T_{N(0)+1})$ and therefore,
\begin{equation}
\label{terzo}
\mathbb{E}[A_{g,v}^{(k,w)}(t,s)1_{\{T_{N(0)+1}\leq
t\}}]=\sum_{\theta=1}^{t}\int_{E}\frac{\dot{q}(g,y,
\theta+v)}{1-H(g,v)} (g)^{k(\theta-1)}\,y^{k}\,
M_{y,0}^{(k,w)}(t-\theta,s-\theta)dy.
\end{equation}
A substitution of $(\ref{primo})$, $(\ref{secondo})$ and
$(\ref{terzo})$ into $(\ref{M})$ concludes the proof.
\begin{flushright}
$\Box $
\end{flushright}
\noindent {\bf{Proof} of Corollary $(\ref{mom1})$
}\\
By convention, for any sequence $\{a(t)\}_{t\in \N}$ we have that
$\prod_{j=t+1}^{t}a(t)=1$.\\
\indent The product growth dividend process for $k\in \{1,2\}$, $w=1$
and $s=t$ is $AP_{g,v}^{(k,1)}(t,t)=A_{g,v}^{(k)}(t)\cdot
A_{G(t),B(t)}^{(1)}(t).
$ According to definition $(\ref{DEF3.1})$ we have 
\begin{equation}
A_{G(t),B(t)}^{(1)}(t;\omega)=\left\{
                \begin{array}{cl}
{\prod_{j=t+1}^{t}G(j;\omega)=1} & \mbox {if $\omega \in
\Omega_{G(t),B(t)}$,}\\
                       0   & \mbox {otherwise,}
                \end{array}
          \right.
\end{equation}
\indent Then $AP_{g,v}^{(k,1)}(t,t)=A_{g,v}^{(k)}(t)$ and
$\mathbb{E}[AP_{g,v}^{(k,1)}(t,t)]=\mathbb{E}[A_{g,v}^{(k)}(t)]$.
Therefore to get the result it is sufficient to take the equation
$(\ref{product moment})$ and to set $w=1$ and $s=t$.
\begin{flushright}
$\Box $
\end{flushright}

\noindent {\bf{Proof} of Theorem $(\ref{teo1})$
}\\
The proof can be accomplished by using similar arguments as those of Theorem $(\ref{theo2})$ and therefore is omitted.
\begin{flushright}
$\Box $
\end{flushright}

\noindent {\bf{Proof} of Theorem $(\ref{theo2})$
}\\
In formula $(\ref{62})$ we established that
\begin{equation}
\begin{aligned}
&
p^{2}(d,g,v)=\sum_{i=1}^{+\infty}\bigg(\frac{\mathbb{E}_{(d,g,v)}[\prod_{j=1}^{t}G^{2}(j)]}{r^{2t}}\bigg)d^{2}\\
&
+2\sum_{t=1}^{+\infty}\sum_{w>t}\bigg(\frac{\mathbb{E}_{(d,g,v)}[\prod_{j=1}^{t}G^{2}(j)\prod_{j=t+1}^{w}G(j)]}{r^{t+w}}\bigg)d^{2}.
\end{aligned}
\end{equation}
Let us consider the term
\begin{equation}
\label{star}
\mathbb{E}_{(d,g,v)}\Big[\prod_{j=1}^{t}G^{2}(j)\prod_{j=t+1}^{w}G(j)\Big]
\end{equation}
\begin{equation*}
\begin{aligned}
&=
\mathbb{E}_{(d,g,v)}\Big[\prod_{j=1}^{t}G^{2}(j)\prod_{j=t+1}^{w-1}G(j)\mathbb{E}_{(D(w-1),G(w-1),B(w-1))}[G(w)]\Big]\\
&=
\mathbb{E}_{(d,g,v)}\Big[\prod_{j=1}^{t}G^{2}(j)\prod_{j=t+1}^{w-1}G(j)M_{G(w-1),B(w-1}^{(1)}(1)\Big].
\end{aligned}
\end{equation*}
From Corollary $(\ref{mom1})$ we know that
\begin{equation*}
\begin{aligned}
M_{G(w-1),B(w-1)}^{(1)}(1)&=\bigg(\frac{1-H(G(w-1),B(w-1)+1)}{1-H(G(w-1),B(w-1))}\bigg)
(G(w-1))\\
& +\int_{E}\frac{\dot{q}(G(w-1),y, B(w-1)+1)}{1-H(G(w-1),B(w-1))} \,y\,
dy,
\end{aligned}
\end{equation*}
then from assumption (A1) we obtain that
$M_{G(w-1),B(w-1)}^{(1)}(1)<\overline{g}$. As a direct consequence, by
iteration we obtain:
\begin{equation}
\label{palla1}
M_{g,v}^{(1)}(t)<(\overline{g})^{t},
\end{equation}
and therefore
\begin{equation}
\mathbb{E}_{(d,g,v)}\Big[\prod_{j=1}^{t}G^{2}(j)\prod_{j=t+1}^{w}G(j)\Big]\leq
\mathbb{E}_{(d,g,v)}\Big[\prod_{j=1}^{t}G^{2}(j)\Big]
(\overline{g})^{w-t}.
\end{equation}
Working analogously on the terms $G^{2}(j),\,j=1,\ldots,t$ we have:
\begin{equation}
\begin{aligned}
&\mathbb{E}_{(d,g,v)}\Big[\prod_{j=1}^{t}G^{2}(j)\prod_{j=t+1}^{w}G(j)\Big]\leq
\mathbb{E}_{(d,g,v)}\Big[\prod_{j=1}^{t-1}G^{2}(j)\mathbb{E}_{(D(t-1),G(t-1),B(t-1))}[G^{2}(t)]\Big]
(\overline{g})^{w-t}\\
&
=\mathbb{E}_{(d,g,v)}\Big[\prod_{j=1}^{t-1}G^{2}(j)M_{G(t-1),B(t-1)}^{(2)}(1)\Big]
(\overline{g})^{w-t}.
\end{aligned}
\end{equation}
From Corollary $(\ref{mom1})$ we know that
\begin{equation*}
\begin{aligned}
M_{G(t-1),B(t-1)}^{(2)}(1)&=\bigg(\frac{1-H(G(t-1),B(t-1)+1)}{1-H(G(t-1),B(t-1))}\bigg)
(G(t-1))^{2}\\
& +\int_{E}\frac{\dot{q}(G(t-1),y, B(t-1)+1)}{1-H(G(t-1),B(t-1))}
\,y^{2}\, dy,
\end{aligned}
\end{equation*}
then from assumption (A2) we obtain that
$M_{G(t-1),B(t-1)}^{(2)}(1)<\overline{g}^{(2)}$. As a direct
consequence, by iteration we obtain
\begin{equation}
\label{casetta}
M_{g,v}^{(2)}(t)<(\overline{g}^{(2)})^{t},
\end{equation}
and in turn we get
\begin{equation}
\label{palla}
\mathbb{E}_{(d,g,v)}\Big[\prod_{j=1}^{t}G^{2}(j)\prod_{j=t+1}^{w}G(j)\Big]\leq
(\overline{g}^{(2)})^{t}\cdot (\overline{g})^{w-t}.
\end{equation}
\indent The inequalities $(\ref{palla})$, $(\ref{palla1})$,
$(\ref{casetta})$ together with the assumptions A1 and A2 imply the
finiteness of prices:
\begin{equation}
\begin{aligned}
&
p^{2}(d,g,v)=\sum_{i=1}^{+\infty}\bigg(\frac{\mathbb{E}_{(d,g,v)}[\prod_{j=1}^{t}G^{2}(j)]}{r^{2t}}\bigg)d^{2}\\
&
+2\sum_{t=1}^{+\infty}\sum_{w>t}\bigg(\frac{\mathbb{E}_{(d,g,v)}[\prod_{j=1}^{t}G^{2}(j)\prod_{j=t+1}^{w}G(j)]}{r^{t+w}}\bigg)d^{2}\\
& \leq
\sum_{i=1}^{+\infty}\bigg(\frac{(\overline{g}^{(2)})^{t}}{r^{2t}}\bigg)d^{2}
+2\sum_{t=1}^{+\infty}\sum_{w>t}\bigg(\frac{(\overline{g}^{(2)})^{t}(\overline{g})^{w-t}}{r^{t+w}}\bigg)d^{2}\leq
+\infty.
\end{aligned}
\end{equation}
Now let us prove the asymptotic condition $(\ref{sette2})$. To this end we define
\begin{equation}
\begin{aligned}
\overline{\psi}_{2}&=\sup_{v\in\N}\sup_{y\in E}\psi_{2}(v,y)\\
& =\sup_{v\in\N}\sup_{y\in E}\Bigg(\sum_{t=1}^{+\infty}\frac{M_{g,v}^{(2)}(t)}{r^{2t}}+2\sum_{t=1}^{+\infty}\sum_{s>0}\frac{M_{g,v}^{(2,1)}(t,s)}{r^{2t+s}}\Bigg),
\end{aligned}
\end{equation}
then it results that 
\begin{equation}
\label{triangolo}
\begin{aligned}
& 0\leq \mathbb{E}_{(d,g,v)}[P^{2}(D(N),G(N),B(N))]\\
& =\mathbb{E}_{(d,g,v)}[\sum_{s=N+1}^{+\infty}\frac{M_{G(t),B(t)}^{(2)}(s)}{r^{2s}}D^{2}(N)+2\sum_{s=N+1}^{+\infty}\sum_{w>s}\frac{M_{G(t),B(t)}^{(2,1)}(N,w)}{r^{N+w}}D^{2}(N)]\\
& \leq \overline{\psi}_{2}\mathbb{E}_{(d,g,v)}[D^{2}(N)].
\end{aligned}
\end{equation}
\indent Consequently we have
\begin{equation}
\label{numero}
\mathbb{E}_{(d,g,v)}[\frac{P^{2}(D(N),G(N),B(N))}{r^{2N}}] \leq \overline{\psi}_{2}\mathbb{E}_{(d,g,v)}[\frac{D^{2}(N)}{r^{2N}}] =\overline{\psi}_{2}\frac{M_{(g,v)}^{(2)}(N)}{r^{2N}}d^{2}.
\end{equation}
 \indent Now take the limit of $(\ref{numero})$ as $N\rightarrow +\infty$ and observe that $(\ref{casetta})$ implies that $\lim_{N\rightarrow +\infty}\frac{M_{(g,v)}^{(2)}(N)}{r^{2N}}=0$, thus
\begin{equation} 
\label{parallele}
\lim_{N\rightarrow
+\infty}\frac{\mathbb{E}_{(d,g,v)}[P^{2}(D(N),G(N),B(N))]}{r^{2N}} = 0.
\end{equation}
\indent It remains to prove the validity of the asymptotic condition $(\ref{sette3})$. Applying the Cauchy-Schwartz inequality we have
\begin{equation}
\begin{aligned}
& \lim_{N\rightarrow
+\infty}\sum_{t=1}^{N}\frac{\mathbb{E}_{(d,g,v)}[D(t)P(N)]}{r^{t+N}}\\
& \leq 
 \lim_{N\rightarrow
+\infty}\sum_{t=1}^{N}\Big(\frac{\mathbb{E}_{(d,g,v)}[D^{2}(t)]}{r^{2t}}\Big)^{\frac{1}{2}}\cdot \Big(\frac{\mathbb{E}_{(d,g,v)}[P^{2}(N)]}{r^{2N}}\Big)^{\frac{1}{2}}\\
&=\lim_{N\rightarrow
+\infty}\Big(\frac{\mathbb{E}_{(d,g,v)}[P^{2}(N)]}{r^{2N}}\Big)^{\frac{1}{2}}\cdot \lim_{N\rightarrow
+\infty}\sum_{t=1}^{N}\Big(\frac{\mathbb{E}_{(d,g,v)}[D^{2}(t)]}{r^{2t}}\Big)^{\frac{1}{2}}.
\end{aligned}
\end{equation}
\indent Now it is sufficient to note that 
\begin{equation}
\label{Arossa}
\lim_{N\rightarrow
+\infty}\Big(\frac{\mathbb{E}_{(d,g,v)}[P^{2}(N)]}{r^{2N}}\Big)^{\frac{1}{2}}=0,
\end{equation}
from $(\ref{parallele})$ and that due to the finiteness of $p^{2}(d,g,v)$ we have that 
\begin{equation}
\label{Brossa}
\lim_{N\rightarrow
+\infty}\sum_{t=1}^{N}\Big(\frac{\mathbb{E}_{(d,g,v)}[D^{2}(t)]}{r^{2t}}\Big)^{\frac{1}{2}}<+\infty .
\end{equation}
Formulas $(\ref{Arossa})$ and $(\ref{Brossa})$ imply
\begin{equation}
\lim_{N\rightarrow
+\infty}\sum_{t=1}^{N}\frac{\mathbb{E}_{(d,g,v)}[D(t)P(N)]}{r^{t+N}}=0.
\end{equation}
\begin{flushright}
$\Box $
\end{flushright}
 
\noindent {\bf{Proof} of Proposition $(\ref{prop3})$
}\\
The proof can be accomplished by using similar arguments as those of Proposition $(\ref{prop4})$ and therefore is omitted.
\begin{flushright}
$\Box $
\end{flushright}

\noindent {\bf{Proof} of Proposition $(\ref{prop4})$
}\\
Formula $(\ref{psi2})$ implies that
\[
\psi_{2}(g,v)=\frac{p^{2}(d,g,v)}{d^{2}}.
\]
\indent We have also seen that the second order moment of the price process at current time $k=0$ is given by
\begin{equation}
\label{addendi}
\begin{aligned}
& p^{2}(d,g,v)=\mathbb{E}_{(d,g,v)}\Big[\big(\frac{D(1)+P(1)}{r}\big)^{2}\Big]\\
& \mathbb{E}_{(d,g,v)}\Big[\frac{D^{2}(1)}{r^{2}}\Big]+\mathbb{E}_{(d,g,v)}\Big[\frac{P^{2}(1)}{r^{2}}\Big]+2\mathbb{E}_{(d,g,v)}\Big[\frac{D(1)+P(1)}{r^{2}}\Big].
\end{aligned}
\end{equation}
Now let us compute these three expectations.
\begin{equation}
\label{ad1}
\begin{aligned}
& \mathbb{E}_{(d,g,v)}\Big[\frac{D^{2}(1)}{r^{2}}\Big]=\frac{1}{r^{2}}\mathbb{E}_{(d,g,v)}[G^{2}(1)D^{2}(0)]=\frac{d^{2}}{r^{2}}\mathbb{E}_{(d,g,v)}[G^{2}(1)]\\
& =\frac{d^{2}}{r^{2}}\Big(\frac{1-H(g,v+1)}{1-H(g,v)}
g^{2}+\int_{E}x^{2}\frac{\dot{q}(g,x, v+1)}{1-H(g,v)}dx,\Big)=:\frac{d^{2}}{r^{2}}E(g,v).
\end{aligned}
\end{equation}
\begin{equation}
\label{ad2}
\begin{aligned}
& \mathbb{E}_{(d,g,v)}\Big[\frac{P^{2}(1)}{r^{2}}\Big]=\frac{1}{r^{2}}\mathbb{E}_{(d,g,v)}[\psi_{2}(G(1),B(1)) D^{2}(1)]=\frac{d^{2}}{r^{2}}\mathbb{E}_{(d,g,v)}[\psi_{2}(G(1),B(1)) G^{2}(1)]\\
& =\frac{d^{2}}{r^{2}}\Big(\frac{1-H(g,v+1)}{1-H(g,v)}
\psi_{2}(g,v+1)g^{2}+\int_{E}\psi_{2}(x,0)x^{2}\frac{\dot{q}(g,x, v+1)}{1-H(g,v)}dx,\Big).
\end{aligned}
\end{equation}
It remains to compute the third expectation:
\begin{equation}
\label{ad3}
\begin{aligned}
& \frac{2}{r^{2}}\mathbb{E}_{(d,g,v)}[D(1)P(1)]=
\frac{2}{r^{2}}\mathbb{E}_{(d,g,v)}[G(1)D(0)P(1)]=\frac{2d}{r^{2}}\mathbb{E}_{(d,g,v)}[G(1)P(1)]\\
& =\frac{2d}{r^{2}}\mathbb{E}_{(d,g,v)}[G(1)\psi_{1}(G(1),B(1)) D(1)]=\frac{2d}{r^{2}}\mathbb{E}_{(d,g,v)}[G(1)\psi_{1}(G(1),B(1)) G(1)D(0)]\\
& =\frac{2d^{2}}{r^{2}}\Big(\frac{1-H(g,v+1)}{1-H(g,v)}
\psi_{1}(g,v+1)g^{2}+\int_{E}\psi_{1}(x,0)x^{2}\frac{\dot{q}(g,x, v+1)}{1-H(g,v)}dx,\Big).
\end{aligned}
\end{equation}
A substitution of $(\ref{ad1})$, $(\ref{ad2})$ and $(\ref{ad3})$ into $(\ref{addendi})$ gives
\begin{equation}
\label{torearrange}
\begin{aligned}
& p^{2}(d,g,v)=\psi_{2}(g,v)d^{2}=\frac{d^{2}}{r^{2}}E^{2}(g,v)+\frac{d^{2}}{r^{2}}\Big(\frac{1-H(g,v+1)}{1-H(g,v)}
\psi_{2}(g,v+1)g^{2}\\
&+\int_{E}\psi_{2}(x,0)x^{2}\frac{\dot{q}(g,x, v+1)}{1-H(g,v)}dx,\Big)+\frac{2d^{2}}{r^{2}}\Big(\frac{1-H(g,v+1)}{1-H(g,v)}
\psi_{1}(g,v+1)g^{2}\\
&+\int_{E}\psi_{1}(x,0)x^{2}\frac{\dot{q}(g,x, v+1)}{1-H(g,v)}dx,\Big).
\end{aligned}
\end{equation}
A simple rearrangement of the terms gives equation $(\ref{sistema2})$
\begin{flushright}
$\Box $
\end{flushright}

\end{document}